\documentclass[dvipdfmx,12pt,a4paper]{article}
\usepackage{amsmath,amssymb,cite,slashed,tabularx}
\usepackage[dvipdfmx]{graphicx}

\allowdisplaybreaks
\makeatletter
\newcommand{\@authornote}[2]{{\def\thefootnote{\fnsymbol{footnote}}\setcounter{footnote}{#1}#2\setcounter{footnote}{0}}}
\newcommand{\authornotemark}[1]{\@authornote#1{\addtocounter{footnote}{-1}\footnotemark}}
\newcommand{\authornotetext}[2]{\@authornote#1{\footnotetext{#2}}}
\makeatother

\begin{document}

\renewcommand{\thefootnote}{\fnsymbol{footnote}} 
\begin{titlepage}

\begin{center}

\hfill KEK-TH-1921\\

\vskip .75in

{\Large\bf
Chargino contributions in light of recent $\epsilon'/\epsilon$
}

\vskip .75in

{\large
  Motoi Endo$^{\rm (a,b)}$, 
  Satoshi Mishima$^{\rm (a)}$,
  Daiki Ueda$^{\rm (b)}$ and
  Kei Yamamoto$^{\rm (a)}$
}

\vskip 0.25in

$^{\rm (a)}${\em 
Theory Center, IPNS, KEK, Tsukuba, Ibaraki 305-0801, Japan}

\vskip 0.1in

$^{\rm (b)}${\em 
The Graduate University of Advanced Studies (Sokendai),\\
Tsukuba, Ibaraki 305-0801, Japan}

\end{center}

\vskip .5in

\begin{abstract}

Recently, the standard model prediction of $\epsilon'/\epsilon$ was improved, and a discrepancy from the experimental results was reported at the $2.9\sigma$ level.
We study the chargino contributions to $Z$ penguin especially with the vacuum stability constraint.
The vacuum decay rate is investigated, and it is shown that the discrepancy can be explained if superparticles are lighter than 4--6\,TeV. 
Correlations with $\mathcal{B}(K_L\to\pi^0\nu\bar\nu)$ and other experimental constraints are also discussed. 

\end{abstract}
\end{titlepage}

\setcounter{page}{1}
\renewcommand{\thefootnote}{\#\arabic{footnote}}
\setcounter{footnote}{0}

\section{Introduction}
\setcounter{equation}{0}
\label{sec:intro}

One of the most sensitive probes of physics beyond the standard model (SM) has been provided by CP-violating observables of flavor-changing neutral currents (FCNCs) in $K$ meson processes. 
The hadron matrix elements of the $K\to\pi\pi$ decay are recently determined with lattice QCD by the RBC-UKQCD collaborations~\cite{Blum:2015ywa}, and the SM prediction of the direct CP violation is obtained as
\begin{align}
 \left(\epsilon'/\epsilon\right)_{\rm SM} = (1.38 \pm 6.90) \times 10^{-4}.~~~[\mbox{RBC-UKQCD}]
\end{align}
The hadronic uncertainties are reduced by the use of CP-conserving data as~\cite{Buras:2015yba}
\begin{align}
 \left(\epsilon'/\epsilon\right)_{\rm SM} = (1.9 \pm 4.5) \times 10^{-4}.~~~[\mbox{Buras et al.}]
 \label{eq:epsSM}
\end{align}
By improving the RG evolutions, one obtains~\cite{Kitahara:2016nld}
\begin{align}
 \left(\epsilon'/\epsilon\right)_{\rm SM} = (0.96 \pm 4.96) \times 10^{-4}.~~~[\mbox{Kitahara et al.}]
 \label{eq:epsSM2}
\end{align}
These SM predictions are lower than the experimental result~\cite{PDG},
\begin{align}
 \left(\epsilon'/\epsilon\right)_{\rm exp} = (16.6 \pm 2.3) \times 10^{-4},
\end{align}
from the NA48~\cite{Batley:2002gn} and KTeV collaborations~\cite{AlaviHarati:2002ye,Worcester:2009qt}.
In particular, Eqs.~\eqref{eq:epsSM} and \eqref{eq:epsSM2} disagree with the experimental data at the $2.9\sigma$ level.

The above discrepancy has been discussed in several new physics models~\cite{Blanke:2015wba,Buras:2015yca,Buras:2015kwd,Buras:2015jaq,Buras:2016dxz} including the supersymmetry (SUSY) models~\cite{Tanimoto:2016yfy,Kitahara:2016otd}.
In SUSY, it has been argued that isospin-violating contributions from gluino box diagrams can be responsible for the discrepancy~\cite{Kitahara:2016otd}.
In this letter, we study chargino $Z$-penguin contributions as an alternative scenario.
They do not decouple even if SUSY particles are heavy as long as a product of the mass insertion (MI) parameters $(\delta_{LR}^u)_{13}^* (\delta_{LR}^u)_{23}$ is fixed~\cite{Colangelo:1998pm,Buras:1999da}.
This feature is attractive once other constraints are considered.
For instance, although CP-violating FCNCs of the $K$ meson are tightly constrained by the indirect CP violation of the $K$ meson or electric dipole moments, SUSY contributions to them decouple in heavy SUSY scenarios. 
Thus, the discrepancy may be explained by the chargino contributions to the $Z$ penguin in $\epsilon'/\epsilon$. 

Among the constraints, one should pay attention that the vacuum stability condition is not relaxed even if SUSY particles are heavy. 
Since the MI parameters, $(\delta_{LR}^u)_{13}$ and $(\delta_{LR}^u)_{23}$, are proportional to scalar trilinear couplings, the chargino $Z$-penguin contributions are constrained by requiring the stability of the electroweak (EW) vacuum.  
In the literature, charge-color breaking (CCB) vacua or potential directions unbounded from below (UFB) have been studied along with $\epsilon'/\epsilon$~\cite{Colangelo:1998pm,Buras:1999da}. 
However, their analyses follow the strategy of Ref.~\cite{Casas:1996de,Casas:1997ze}, and the vacuum decay rate has not been examined.
In this letter, the vacuum decay will be studied, and we will discuss whether the current discrepancy of $\epsilon'/\epsilon$ is explained by the chargino $Z$-penguin contributions.

\section{Signals and constraints}
\setcounter{equation}{0}

\subsection{Notations}
We basically follow the definition of SUSY Les Houches accord (SLHA) to describe the SUSY Lagrangian~\cite{Skands:2003cj,Allanach:2008qq}. 
The up-type squarks and charginos appear in the chargino contributions to the flavor-violating $Z$-boson couplings of the down-type quarks. 
In terms of the squark fields, $\Phi_u=(\tilde u_L,\tilde c_L,\tilde t_L,\tilde u_R,\tilde c_R,\tilde t_R)^T$, the up-type squark mass matrix is described as
\begin{align}
 \mathcal{M}_{\tilde u}^2 = 
 \begin{pmatrix}
 m_{\tilde Q}^2 + m_u^2 + \cos 2\beta\, m_Z^2 \left(\frac{1}{2}-\frac{2}{3}\sin^2\theta_W\right) & 
 \frac{v_2}{\sqrt{2}}T_U^*-\mu m_u \cot\beta \\
 \frac{v_2}{\sqrt{2}}T_U^T-\mu^* m_u \cot\beta & 
 m_{\tilde U}^{2T} + m_u^2 + \frac{2}{3} \cos 2\beta\, m_Z^2 \sin^2\theta_W
 \end{pmatrix}.
\end{align}
It is diagonalized by a unitary matrix $\mathcal{R}^u$,
\begin{align}
 \mathcal{R}^u \mathcal{M}_{\tilde u}^2 \mathcal{R}^{u\dagger} = {\rm diag}(m_{\tilde u_i}^2).
\end{align}
In this letter, the soft mass parameters are set in the superCKM basis, where the Yukawa matrices are diagonalized.
Although the soft SUSY-breaking masses, $m_{\tilde Q}^2$ and $m_{\tilde U}^2$, generally have flavor off-diagonal components, they are irrelevant for the current discrepancy of $\epsilon'/\epsilon$, because SUSY contributions to the $Z$ penguin are enhanced when the SU(2)$_L$ symmetry is broken, as will be mentioned in the next section.
A significant contribution is provided by flavor mixings in the trilinear scalar coupling $T_U$, which is also expressed by the MI parameters,
\begin{align}
 (\delta^u_{LR})_{ij} = \frac{\frac{v_2}{\sqrt{2}} (T_U)_{ij}^*}{m_{\tilde q}^2},~~~
 (\delta^u_{RL})_{ij} = \frac{\frac{v_2}{\sqrt{2}} (T_U)_{ji}}{m_{\tilde q}^2}.
 \label{eq:defMIA}
\end{align}
Here, $m_{\tilde q}$ is a squark mass.
It is noted that $(T_U)_{ij}$ and $(\delta^u_{LR})_{ij}$ are complex parameters, and $(\delta_{LR}^u)_{ij} = (\delta_{RL}^u)_{ji}^*$ is satisfied.

The chargino mass matrix is given by
\begin{align}
 \mathcal{M}_{\tilde\psi^+} = 
 \begin{pmatrix}
 M_2 & \sqrt{2} m_W \sin\beta \\
 \sqrt{2} m_W \cos\beta & \mu
 \end{pmatrix},
\end{align}
which is diagonalized by two unitary matrices $\mathcal{U}$ and $\mathcal{V}$ as
\begin{align}
 \mathcal{U}^* \mathcal{M}_{\tilde\psi^+} \mathcal{V}^\dagger = {\rm diag}(m_{\tilde\chi^+_i}).
\end{align}

\subsection{$K$ meson observables}

\begin{figure}
\begin{center}
\includegraphics[scale=1]{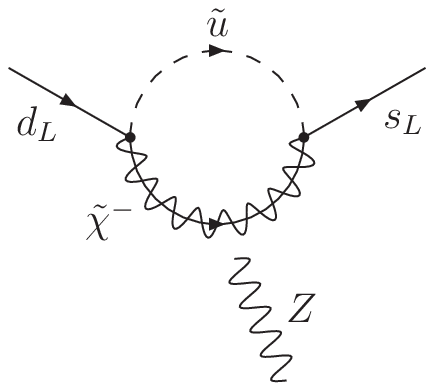}\hspace{10mm}
\includegraphics[scale=1]{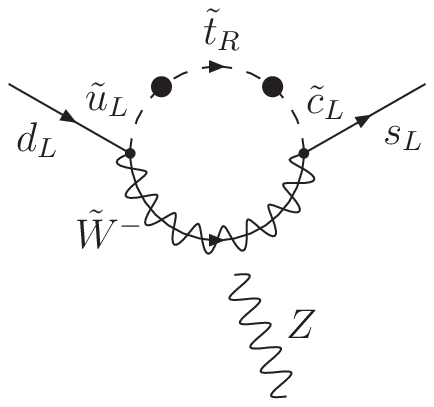}
\end{center}
\caption{
 Chargino contributions to $Z$ penguin are shown. 
 SUSY particles are denoted in the mass eigenstate (left), while they are in the MI approximation (right).
 The MI parameters are inserted at the black dots on the squark line in the right plot.
}
\label{fig:diagram}
\end{figure}

Chargino contributions to the $Z$-penguin diagrams are studied in this letter.
They are described by the flavor-violating $Z$-boson vertex, 
\begin{align}
 \mathcal{L}_{\rm eff} = 
 \frac{-g^3}{8\pi^2\cos\theta_W}
 Z_{ds}\,\bar s_L \gamma_\mu d_L Z^\mu + h.c.\,.
\end{align}
The coupling includes SM and SUSY contributions,
\begin{align}
 Z_{ds} = Z_{ds}^{\rm (SM)} + Z_{ds}^{\rm (SUSY)}.
\end{align}
The SM term is dominated by the top-quark loop contributions (see e.g.,~\cite{Buras:1999da}),
\begin{align}
 Z_{ds}^{\rm (SM)} = \lambda_t C_{\rm SM}(x_t),~~~
 C_{\rm SM}(x) = 
 \frac{x}{8} \left[\frac{x-6}{x-1} + \frac{3x+2}{(x-1)^2}\ln x\right],
\end{align}
where $\lambda_t = V_{ts}^*V_{td}$ and $x_t = m_t^2/m_W^2$.
On the other hand, the SUSY contribution is given by the up-type squark-chargino diagrams (left in Fig.~\ref{fig:diagram}) as (c.f., Ref.~\cite{Hofer:2010ee})
\begin{align}
 \label{eq:ZdsSUSY}
 Z_{ds}^{\rm (SUSY)} &=
 \frac{1}{8g^2} \Gamma^L_{r2n} \Gamma^{L*}_{s1m}
 \bigg( \delta_{mn} C_2(m_{\tilde \chi_m}^2,m_{\tilde u_s}^2,m_{\tilde u_r}^2)
 \sum_{k=1}^3 \mathcal{R}_{rk}^u \mathcal{R}_{sk}^{u*} \\
 &~~~
 + \delta_{sr}\left[
 2 C_0(m_{\tilde u_s}^2,m_{\tilde \chi_m}^2,m_{\tilde \chi_n}^2)
 m_{\tilde \chi_m}m_{\tilde \chi_n}
 \mathcal{U}_{n1}\mathcal{U}_{m1}^* - 
 C_2(m_{\tilde u_s}^2,m_{\tilde \chi_m}^2,m_{\tilde \chi_n}^2)
 \mathcal{V}_{m1}\mathcal{V}_{n1}^* \right] \bigg). \notag
\end{align}
Here, the chargino-quark-squark coupling is defined as
\begin{align}
 \Gamma^L_{rim} = 
 (g \mathcal{R}_{rk}^{u*} \mathcal{V}_{m1} - \mathcal{R}_{r,k+3}^{u*} \mathcal{V}_{m2} Y_{u_k}) V_{ki}^*,
\end{align}
where $Y_{u_k}$ is the up-type quark Yukawa coupling, and $V_{ki}$ the CKM matrix. The loop functions are 
\begin{align}
 C_0(x,y,z) &= -\frac{y}{(x-y)(z-y)}\ln \frac{y}{x} + (y\leftrightarrow z), \\
 C_2(x,y,z) &= -\frac{y^2}{(x-y)(z-y)}\ln \frac{y}{x} + (y\leftrightarrow z) \notag \\
 &~~~ 
 + \frac{2}{4-d} + \ln 4\pi -\gamma_E + \ln\frac{\mu^2}{x} + 1.
\end{align}
The second line in $C_2$ is from a regularization.
It is noticed that $Z_{ds}^{\rm (SUSY)}$ is independent of its constant because of the unitarity of the mixing matrix of squarks. 

It is instructive to represent the SUSY contribution in terms of the MI approximation.
This is achieved by expanding Eq.~\eqref{eq:ZdsSUSY} in terms of $\mathcal{O}(m_W^2/m_{\tilde q}^2)$ and $(\delta_{LR}^u)_{ij}$.
Focusing on the MI parameters, $(\delta_{LR}^u)_{13}$ and $(\delta_{LR}^u)_{23}$, one obtains
\begin{align}
 Z_{ds}^{\rm (SUSY)} \simeq (\delta_{LR}^u)_{13}^* (\delta_{LR}^u)_{23} H_0(x_{\tilde q\tilde W}),
 \label{eq:SUSYMIA}
\end{align}
which reproduces the result in Ref.~\cite{Colangelo:1998pm,Buras:1999da}.
The loop function is defined as
\begin{align}
 H_0(x) = -\frac{x(x^3-6x^2+3x+2+6x\ln x)}{48(1-x)^4},
\end{align}
with $x_{\tilde q\tilde W} = m_{\tilde q}^2/m_{\tilde W}^2$.
The squark masses are supposed to be degenerate, $m_{\tilde Q} = m_{\tilde U} \equiv m_{\tilde q}$.
Since it depends on a ratio of the SUSY masses, $Z_{ds}^{\rm (SUSY)}$ is not suppressed by heavy SUSY particles as long as $(\delta_{LR}^u)_{13}^* (\delta_{LR}^u)_{23}$ is fixed.
It also does not vanish at $q^2=0$, where $q^2$ is the momentum transfer of the $Z$ boson.
These features are guaranteed by the SU(2)$_L$ breaking, which is provided by $(\delta_{LR}^u)_{13}$ and $(\delta_{LR}^u)_{23}$ in Eq.~\eqref{eq:SUSYMIA}.
SUSY contributions including other MI parameters are suppressed e.g., by higher orders of $\mathcal{O}(m_W^2/m_{\tilde q}^2)$.
On the other hand, Eq.~\eqref{eq:SUSYMIA} corresponds to diagrams involving the left-handed sup and scharm, the right-handed stop, and the Wino in the loop (right in Fig.~\ref{fig:diagram}). 
Hence, the loop function depends only on their masses, and the model parameters relevant for $Z_{ds}^{\rm (SUSY)}$ are
\begin{align}
 m_{\tilde u_L},~
 m_{\tilde c_L},~
 m_{\tilde t_R},~
 m_{\tilde W},~
 (\delta_{LR}^u)_{13},~
 (\delta_{LR}^u)_{23}.
\end{align}
In addition, $m_{\tilde t_L}$ and $A_t$ could contribute if the left-right mixing of the stop is necessary. 
Also, $\tan\beta$ is irrelevant unless it is small.
Higgsino contributions are suppressed by tiny Yukawa couplings.
In this letter, the gluino mass is assumed to be very large so that gluino contributions to $\epsilon'/\epsilon$ are suppressed.
SUSY box contributions to $\epsilon'/\epsilon$ are neglected because they are small.

The theoretical value of $\epsilon'/\epsilon$ is composed by the SM and SUSY contributions,
\begin{align}
 \left(\epsilon'/\epsilon\right) = 
 \left(\epsilon'/\epsilon\right)_{\rm SM} +
 \left(\epsilon'/\epsilon\right)_{\rm SUSY}.
\end{align}
As mentioned in Sec.~\ref{sec:intro}, the SM one has been improved recently~\cite{Blum:2015ywa,Buras:2015yba,Kitahara:2016nld}.
The discrepancy is estimated as
\begin{align}
 \Delta\left(\epsilon'/\epsilon\right) = 
 \left\{
 \begin{array}{l}
 (15.2 \pm 7.3) \times 10^{-4},~~~(2.1\sigma)~~~[\mbox{RBC-UKQCD}] \\
 (14.7 \pm 5.1) \times 10^{-4},~~~(2.9\sigma)~~~[\mbox{Buras et al.}] \\
 (15.6 \pm 5.5) \times 10^{-4},~~~(2.9\sigma)~~~[\mbox{Kitahara et al.}]
 \end{array}
 \right.
\end{align}
where $\Delta\left(\epsilon'/\epsilon\right) \equiv \left(\epsilon'/\epsilon\right)_{\rm exp}-\left(\epsilon'/\epsilon\right)_{\rm SM}$, and the errors are summed in quadrature. 
We examine whether $\left(\epsilon'/\epsilon\right)_{\rm SUSY}$ saturates these gaps.
The $Z$ penguin contribution is expressed as~\cite{Buras:1999da,Buras:2000qz},
\begin{align}
 \left(\epsilon'/\epsilon\right)_Z = (P_X + P_Y + P_Z)\,
 {\rm Im}\,Z_{ds}\,.
 \label{eq:epsZ}
\end{align}
Here, $P_X$, $P_Y$ and $P_Z$ assemble the information below the weak scale such as hadron matrix elements and QCD corrections. 
Their numerical results are~\cite{Buras:2015yba},
\begin{align}
 P_X + P_Y + P_Z = 1.52 + 0.12 R_6 - 13.65 R_8,
\end{align}
where $R_6$ and $R_8$ are defined as
\begin{align}
 R_6 = B_6^{(1/2)}(m_c) \left[ \frac{114.54\,{\rm MeV}}{m_s(m_c)+m_d(m_c)} \right]^2,~~~
 R_8 = B_8^{(3/2)}(m_c) \left[ \frac{114.54\,{\rm MeV}}{m_s(m_c)+m_d(m_c)} \right]^2,
\end{align}
and hadron matrix elements are~\cite{Blum:2015ywa,Buras:2015yba}
\begin{align}
 B_6^{(1/2)}(m_c) = 0.57 \pm 0.19,~~~
 B_8^{(3/2)}(m_c) = 0.76 \pm 0.05.
\end{align}
Therefore, $P_X+P_Y+P_Z$ is negative.

The SUSY contribution \eqref{eq:ZdsSUSY} is evaluated at the SUSY scale, which is higher than the weak scale.
Renormalization group (RG) corrections between the SUSY and weak scales are subleading. 
Those below the weak scale are included in $P_i$.
Above the weak scale, the SU(2)$_L$ symmetry is restored, and the effective $Z$-boson vertex is described by the dimension-6 operators, $(H^\dagger i\overleftrightarrow{D}_\mu H)(\bar q'\gamma^\mu q)$ and $(H^\dagger i\overleftrightarrow{D}_\mu^I H)(\bar q'\tau^I\gamma^\mu q)$.
Anomalous dimensions of their RG equations are not large~\cite{Jenkins:2013zja,Jenkins:2013wua,Alonso:2013hga}. 
This is not the case of Ref.~\cite{Kitahara:2016nld}, where the effective operators of $s \to dq\bar q$ are generated at the SUSY scale.

The flavor-changing $Z$-boson coupling also contributes to $K_L \to \pi^0\nu\bar\nu$.
The branching ratio is expressed as~\cite{Colangelo:1998pm,Buras:2015yca}
\begin{align}
 \mathcal{B}(K_L\to\pi^0\nu\bar\nu) = 
 \kappa_L \left[ 
 \frac{{\rm Im}\big(\lambda_t X^{\rm (SM)}+Z_{ds}^{\rm (SUSY)}\big)}{\lambda^5} 
 \right]^2,
\end{align}
where $\kappa_L = (2.231 \pm 0.013) \cdot 10^{-10}(\lambda/0.225)^8$, $X^{\rm (SM)} = 1.481 \pm 0.009$ and $\lambda$ is the Wolfenstein parameter.
The SM prediction is about $2.8\times10^{-11}$~\cite{Buras:2015jaq}.
Compared with Eq.~\eqref{eq:epsZ}, it is noticed that the SUSY contribution to $\mathcal{B}(K_L\to\pi^0\nu\bar\nu)$ has a negative correlation with that to $\epsilon'/\epsilon$ as long as it is dominated by the chargino $Z$-penguin contribution (c.f., Ref.~\cite{Buras:2015yca}).
Although $K^+ \to \pi^+\nu\bar\nu$ includes a similar contribution, its effect is weak.

\subsection{Vacuum stability}

According to Eq.~\eqref{eq:SUSYMIA}, large $\epsilon'/\epsilon$ is achieved when $\tilde u_L$ and $\tilde c_L$ have a large mixing with $\tilde t_R$. 
The left-right mixing is proportional to the scalar trilinear coupling $(T_U)_{ij}$. 
Large flavor-violating trilinear couplings may generate instabilities of the EW vacuum~\cite{Park:2010wf}.
Requiring that the lifetime of the EW vacuum is longer than the present age of the universe, the trilinear couplings, or equivalently $(\delta_{LR}^u)_{13}$ and $(\delta_{LR}^u)_{23}$, are constrained.

The vacuum decay rate per unit volume is expressed as
\begin{align}
 \Gamma/V = A \exp (-S_E).
\end{align}
In this letter, $S_E$ is estimated at the semi-classical level, which is called the bounce action~\cite{Coleman:1977py} and calculated by {\tt CosmoTransition} 2.0a2~\cite{Wainwright:2011kj}.
The prefactor $A$ is not determined at this level; higher-order calculations are needed for determining $A$~\cite{Callan:1977pt}.
We adopt an order-of-estimation analysis. 
Since typical energy scales are the EW and SUSY ones, we take $A \sim (100\,{\rm GeV})^4$ or $(10\,{\rm TeV})^4$.
The lifetime of the EW vacuum is longer than the age of the universe if the bounce action satisfies
\begin{align}
  S_E \gtrsim 400.
\end{align}
Thermal effects are neglected in this letter.

The bounce action potentially involves $\mathcal{O}(10\%)$ uncertainties due to renormalization scale dependences of the model parameters. 
They are improved by taking radiative corrections into account~\cite{Endo:2015ixx}.
However, they are neglected in this letter for simplicity; calculations of the radiative corrections are complicated and will be studied elsewhere. 

The bounce action is calculated once the scalar potential is given. 
In the superCKM basis, the relevant part of the potential is given by
\begin{align}
 V =& 
 \frac{1}{2} m_{11}^2 \, h_d^2 
 + \frac{1}{2} m_{22}^2 \, h_u^2 
 - m_{12}^2 \, h_d h_u 
 + \frac{1}{2} m_{\tilde Q_i}^2 \,\tilde u_{iL}^2 
 + \frac{1}{2} m_{\tilde Q_3}^2 \,\tilde t_L^2 
 + \frac{1}{2} m_{\tilde U_3}^2 \,\tilde t_R^2 
 \notag \\ & 
 + \frac{1}{\sqrt{2}} \left[(T_U)_{33} h_u - y_t \mu h_d\right]\tilde t_L \tilde t_R 
 + \frac{1}{\sqrt{2}} (T_U)_{i3} h_u \tilde u_{iL} \tilde t_R 
 + \frac{1}{4} y_t^2 (\tilde t_L^2 \tilde t_R^2 + \tilde t_L^2 h_u^2 + \tilde t_R^2 h_u^2) 
 \notag \\ & 
 + \frac{1}{24} g_3^2 (\tilde u_{iL}^2 + \tilde t_L^2 - \tilde t_R^2)^2
 + \frac{1}{32} g^2 (h_u^2 - h_d^2 - \tilde u_{iL}^2 - \tilde t_L^2)^2
 \notag \\ & 
 + \frac{1}{32} g_Y^2 \left(h_u^2 - h_d^2 + \frac{1}{3}\tilde u_{iL}^2 + \frac{1}{3}\tilde t_L^2 - \frac{4}{3}\tilde t_R^2 \right)^2,
 \label{eq:potential}
\end{align}
where $h_u$, $h_d$, $\tilde u_{iL}$, $\tilde t_R$ and $\tilde t_L$ are real scalar fields. 
Here, $\tilde u_{iL}$ denotes the left-handed sup or scharm ($i=1,2$).
Terms including $y_u$ or $y_c$ are neglected, while mixings of $h_u$--$h_d$ or $\tilde t_R$--$\tilde t_L$ are kept included. 
The coefficients in the Higgs sector are expressed as
\begin{align}
 m_{11}^2 = m_A^2 \sin^2\beta - \frac{1}{2} m_Z^2 \cos 2\beta,~
 m_{22}^2 = m_A^2 \cos^2\beta + \frac{1}{2} m_Z^2 \cos 2\beta,~
 m_{12}^2 = \frac{1}{2} m_A^2 \sin 2\beta.
\end{align}

In general, $(T_U)_{13}$ and $(T_U)_{23}$ have complex phases.
They can be rephased out in the potential \eqref{eq:potential} before taking real parts of the fields, and the model parameters are set to be real.
Thus, the vacuum stability conditions provide upper bounds on the magnitude of $(T_U)_{ij}$.

Two trilinear couplings $(T_U)_{13}$ and $(T_U)_{23}$ generate two CCB vacua.
In the calculation of $S_E$, one CCB vacuum does not affect another.
The bounce action is a solution of the Euclidean equation of motion. 
A semi-classical path belonging to one CCB vacuum is hardly affected by another.
Therefore, the bounce actions are calculated for $(T_U)_{13}$ and $(T_U)_{23}$, separately.

The trilinear coupling is composed by $h_u$, $\tilde u_{iL}$ and $\tilde t_R$. 
In the limit when heavy Higgs bosons are decoupled and the stop left-right mixing is negligible, $h_u$ becomes close to the SM-like Higgs boson $H$, and $\tilde t_L$ does not contribute to the vacuum decay rate.
Then, the scalar potential is expressed by $H$, $\tilde u_{iL}$ and $\tilde t_R$ as 
\begin{align}
 V =& 
 - \frac{1}{4} m_Z^2 \cos^2 2\beta\, H^2 
 + \frac{1}{2} m_{\tilde Q_i}^2 \,\tilde u_{iL}^2 
 + \frac{1}{2} m_{\tilde U_3}^2 \,\tilde t_R^2 
 + \frac{1}{\sqrt{2}} (T_U)_{i3} \sin\beta\, H \tilde u_{iL} \tilde t_R 
 + \frac{1}{4} y_t^2 \sin^2\beta\, H^2 \tilde t_R^2 
 \notag \\ & 
 + \frac{1}{24} g_3^2 (\tilde u_{iL}^2 - \tilde t_R^2)^2
 + \frac{1}{32} g^2 (H^2 \cos2\beta + \tilde u_{iL}^2)^2
 + \frac{1}{32} g_Y^2 \left(H^2 \cos2\beta - \frac{1}{3}\tilde u_{iL}^2 + \frac{4}{3}\tilde t_R^2 \right)^2.
\end{align}
In the potential, the mass of the SM-like Higgs boson is lower than $125\,{\rm GeV}$, which is cured by radiative corrections to the Higgs potential. 
Including such corrections to the vacuum decay rate is beyond the scope of the analysis in this letter.

\section{Results}
\setcounter{equation}{0}

\begin{figure}
\begin{center}
\includegraphics[scale=1]{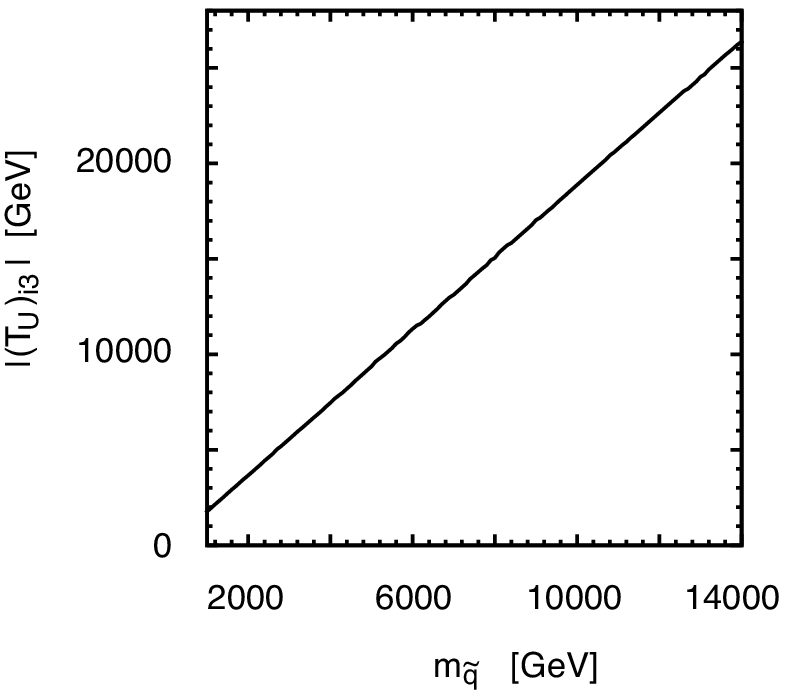}\hspace{5mm}
\includegraphics[scale=1]{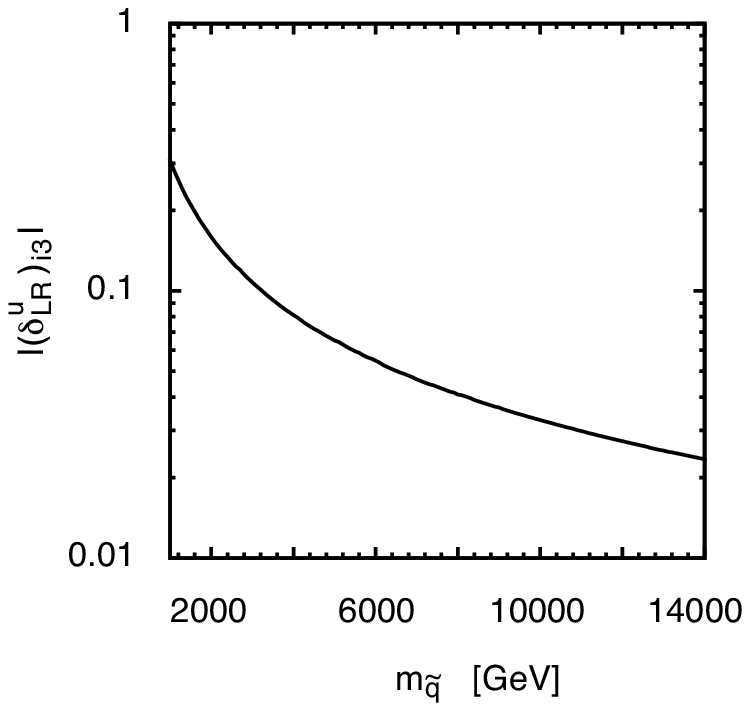}
\end{center}
\caption{
  Vacuum stability constraint on $|(T_U)_{i3}|$ (left) and $|(\delta_{LR}^u)_{i3}|$ (right) for $i=1,2$ as a function of $m_{\tilde q}$. 
  Here, $m_{\tilde q} \equiv m_{\tilde Q_i} = m_{\tilde U_3}$ and $\tan\beta = 50$.
  It is assumed that the heavy Higgs bosons are decoupled and the stop left-right mixing is neglected.
}
\label{fig:vac}
\end{figure}

We discuss whether the current discrepancy of $\epsilon'/\epsilon$ is explained by the chargino $Z$-penguin contributions with satisfying the constraints especially from the vacuum stability condition.
First, the vacuum decay rate is estimated to derive an upper bound on the size of $(T_U)_{i3}$ by requiring $S_E \gtrsim 400$.
In the left plot of Fig.~\ref{fig:vac}, the bound is shown as a function of $m_{\tilde q} \equiv m_{\tilde Q_i} = m_{\tilde U_3}$. 
Here and hereafter, it is assumed that the heavy Higgs bosons are decoupled and the left-right mixing of stops is neglected. 
The result is insensitive to $\tan\beta$ as long as it is large.
In the right plot, the result is interpreted into the bound of $(\delta_{LR}^u)_{i3}$.
Due to the relation \eqref{eq:defMIA}, the limit becomes severer as the SUSY scale increases. 
Therefore, the SUSY contributions to $\epsilon'/\epsilon$ decrease according to Eq.~\eqref{eq:SUSYMIA}.

\begin{figure}[t]
\begin{center}
\includegraphics[scale=1]{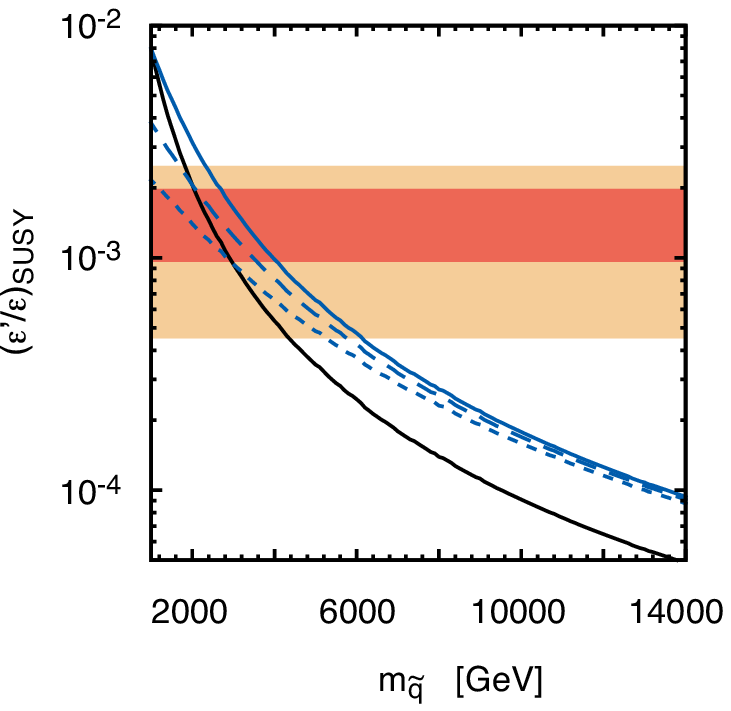}\hspace{5mm}
\includegraphics[scale=1]{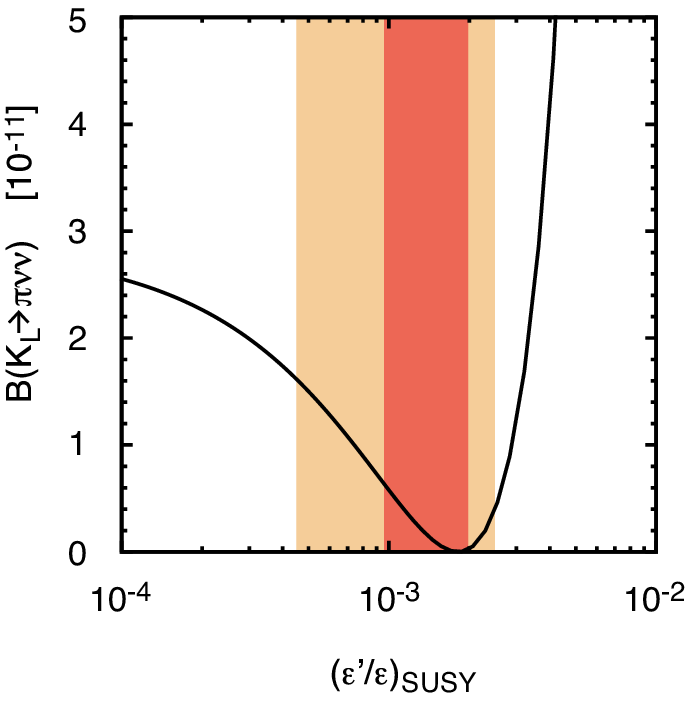}
\end{center}
\caption{
 $\left( \epsilon'/\epsilon \right)_{\rm SUSY}$ is shown as a function of $m_{\tilde q}$ (left). 
 Here, $m_{\tilde q} \equiv m_{\tilde Q_i} = m_{\tilde U_3}$, $\tan\beta = 50$ and $|(T_U)_{13}| = |(T_U)_{23}|$ at $S_E = 400$.
 The CP-violating phase is maximal.
 The Wino mass $m_{\tilde W}$ is 1, 2, 3\,TeV for the blue solid, dashed and dotted lines, respectively, while it is equal to $m_{\tilde q}$ on the black line.
 On the red (orange) region, $\Delta\left(\epsilon'/\epsilon\right)$ is saturated at the $1\sigma$ ($2\sigma$) level. 
 The SM value follows Ref.~\cite{Buras:2015yba}.
 Right: correlation between $\mathcal{B}(K_L\to\pi^0\nu\bar\nu)$ and $\left( \epsilon'/\epsilon \right)_{\rm SUSY}$ is shown. 
}
\label{fig:obs}
\end{figure}

In the left plot of Fig.~\ref{fig:obs}, the SUSY contributions to $\epsilon'/\epsilon$ are shown as a function of $m_{\tilde q}$. 
Here, $|(T_U)_{i3}|$ is set at $S_E = 400$, and $|(T_U)_{13}| = |(T_U)_{23}|$ is assumed.
The CP-violating phase is taken to be maximal.
In addition to the model parameters that determine the vacuum decay rate, there is a degree of freedom in choosing $m_{\tilde W}$ (see Eq.~\eqref{eq:ZdsSUSY}).
In the figure, $m_{\tilde W}$ is set to be 1, 2, 3\,TeV and $m_{\tilde q}$ as reference cases. 
The result is insensitive to the other model parameters.
It is found that the current discrepancy of $\epsilon'/\epsilon$ can be explained; the SUSY scale can be as large as 4--6\,TeV, depending on the choice of $m_{\tilde W}$.

So far, $m_{\tilde Q_i} = m_{\tilde U_3}$ and $|(T_U)_{13}| = |(T_U)_{23}|$ are supposed. 
If we set $m_{\tilde Q_i} \neq m_{\tilde U_3}$ and/or $|(T_U)_{13}| \neq |(T_U)_{23}|$, the SUSY contributions to $\epsilon'/\epsilon$ become smaller at $S_E = 400$.

In the right plot of Fig.~\ref{fig:obs}, correlation between $\mathcal{B}(K_L\to\pi^0\nu\bar\nu)$ and $\left( \epsilon'/\epsilon \right)_{\rm SUSY}$ is displayed.
As mentioned in the previous section, $\mathcal{B}(K_L\to\pi^0\nu\bar\nu)$ decreases as $\epsilon'/\epsilon$ increases unless $\epsilon'/\epsilon$ is very large. 
(When $\epsilon'/\epsilon$ is huge, the SUSY contribution is larger than the SM one for $K_L\to\pi^0\nu\bar\nu$.)
The current discrepancy implies that $\mathcal{B}(K_L\to\pi^0\nu\bar\nu)$ is predicted to be less than 60\% of the SM prediction.
In future, the KOTO experiment may measure the branching ratio at the 10\% level of the SM value~\cite{Shiomi:2014sfa,KOTO}.

Some parameter regions are constrained by other observables. 
Those in $m_{\tilde q} \lesssim 1$--$2\,{\rm TeV}$ are excluded by $\epsilon_K$.
The constraint is given by the chargino box contribution~\cite{Colangelo:1998pm} and relaxed as $m_{\tilde q}$ increases. 
Double penguin contributions using the flavor-changing $Z$-boson coupling~\cite{Buras:2015jaq} are weaker.
A weaker bound is obtained from $\Delta m_d$. 
It changes mainly through box diagrams with $(\delta_{LR}^u)_{13}$~\cite{Colangelo:1998pm}, which decouple as SUSY particles become heavier.
The MI parameter $(\delta_{LR}^u)_{23}$ generates contributions to $\mathcal{B}(b\to s\gamma)$.
Since the dominant contribution is from Higgsino-like chargino diagrams, its effect is sufficiently small if Higgsinos are heavy without suppressing the contribution to $\epsilon'/\epsilon$.
Electric dipole moments are sensitive probes of the CP violations. 
However, contributions with $(T_U)_{13}$ or $(T_U)_{23}$ (see e.g., Ref.~\cite{Endo:2003te}) are smaller than the current experimental limits if the squarks are heavier than 1\,TeV.
Finally, one might obtain a stringent constraint from RG analyses~\cite{Buras:1999da}.
However, they depend on models, and we simply neglect them to keep the discussion as model-independent as possible.

\section{Conclusion}
\setcounter{equation}{0}

The recent analyses of the SM prediction of $\epsilon'/\epsilon$ have reported a discrepancy from the experimental value.
The significance is about the $2.9\sigma$ level.
We studied whether it is explained by the chargino $Z$-penguin contributions.
They are constrained by the vacuum stability condition, and it is found that the SUSY contributions can bridge the discrepancy if the SUSY masses are smaller than 4--6\,TeV. 

The chargino $Z$ penguin also contributes to $\mathcal{B}(K_L\to\pi^0\nu\bar\nu)$.
The current discrepancy of $\epsilon'/\epsilon$ implies that $\mathcal{B}(K_L\to\pi^0\nu\bar\nu)$ is about less than 60\% of the SM prediction.
In future, the KOTO experiment may measure the branching ratio at the 10\% level of the SM value~\cite{Shiomi:2014sfa,KOTO}.
On the other hand, other experimental constraints exclude models only when the SUSY particles are lighter than 1--2\,TeV.

The SM predictions of $\epsilon'/\epsilon$ are expected to be improved in the near future.
If the discrepancy would be confirmed, the chargino contributions could provide an attractive solution.

\vspace{1em}
\noindent {\it Acknowledgements}: 
We thank Toru Goto for helpful discussions.
This work is supported by JSPS KAKENHI No.~16K17681 (M.E.) and 16H03991 (M.E.).


\end{document}